# Observation of the quantum valley Hall state in ballistic graphene superlattices

Katsuyoshi Komatsu[1†*], Yoshifumi Morita[2], Eiichiro Watanabe[3], Daiju Tsuya[3], Kenji Watanabe[4], Takashi Taniguchi[4], and Satoshi Moriyama[1*]

[1] International Center for Materials Nanoarchitectonics, National Institute for Materials Science (NIMS), Tsukuba, Ibaraki 305-0044, Japan.

[2] Faculty of Engineering, Gunma University, Kiryu, Gunma 376-8515, Japan.

[3] Nanofabrication Platform, NIMS, Tsukuba, Ibaraki 305-0047, Japan.

[4] Research Center for Functional Materials, NIMS, Tsukuba, Ibaraki 305-0044, Japan.

*Correspondence to: katsuyoshi1.komatsu@toshiba.co.jp, MORIYAMA.Satoshi@nims.go.jp

†Current address: Institute of Memory Technology Research & Development, Toshiba Memory Corporation, Yokkaichi, Mie 512-8550, Japan.

**Abstract:**

In graphene superlattices, bulk topological currents can lead to long-range charge-neutral flow and non-local resistance near Dirac points. A ballistic version of these phenomena has never been explored. Here, we report transport properties of ballistic graphene superlattices. This allows us to study and exploit giant non-local resistances with a large valley Hall angle without a magnetic field. In the low-temperature regime, a crossover occurs toward a new state of matter, referred to as a quantum valley Hall state (qVHS), which is an analog of the quantum Hall state without a magnetic field. Furthermore, a non-local resistance plateau, implying rigidity of the qVHS, emerges as a function of magnetic field, and the collapse of this plateau is observed, which is considered as a manifestation of valley/pseudospin magnetism.

**Introduction:**

The Berry phase (*1*) is considered an important tool for developing spintronics and valleytronics (*2, 3*) because it allows control over spin/valley degrees of freedom instead of charge degree of freedom. A particularly challenging problem is the engineering of the Berry phase through the bulk topological current, which is a descendant of the quantum Hall effect (QHE) (*4-6*) but can occur even in the absence of a magnetic field. Recent discoveries of the quantum anomalous Hall effect (7), which breaks time-reversal symmetry, and quantum spin Hall/topological insulator (*8*) are also variants of the QHE in terms of topology. Valley current in graphene (*9-13*) offers promise for realizing the bulk topological current in a (2+1)-dimensional Dirac fermion without a magnetic field (*6*). Furthermore, valleytronics (*2*) are expected to fulfill the requirements of the next generation of electronics devices, such as the valley field effect transistor (*11*, *14*).

In monolayer graphene, which exhibits a Dirac-type relativistic energy dispersion (i.e., a Dirac cone), the broken inversion symmetry can be induced by placing graphene on a hexagonal boron nitride (hBN) substrate with a precise alignment angle near zero degrees, whereby a long-length moiré pattern appears between the graphene and the hBN lattice because of the 1.8% lattice mismatch between the graphene and hBN (*15-17*). The moiré superlattice also induces a fractal energy spectrum under a magnetic field, which is known as Hofstadter's butterfly (*15-18*). In these systems with broken inversion symmetry, the valley current can be generated and detected via the valley Hall effect (VHE), even in the absence of a magnetic field: the VHE is induced by the accumulated Berry curvature near the apex of the Dirac cone (i.e., the hot spot) and is associated with the topological current (*9-13*). The valley Hall current has not yet been explored near the ballistic regime, and the longitudinal conductivity has always dominated the valley Hall conductivity, implying a small valley Hall angle that veils the nature of the VHE. Observing a coherent neutral long-ranged valley current by means of an all-electrical method requires non-local resistance measurements from a geometry smaller than or comparable to the mean free path (*11*).

**Results and Discussion:**

Toward this objective, we fabricated hBN/graphene/hBN heterostructures with one-dimensional Cr/Au contacts, as shown in Fig. 1(A) and (B). Sharp increases in the longitudinal resistivity ($\rho_{xx}$) at backgate voltages $V_g$ of approximately 0 V and -21 V correspond to a Dirac point (DP) and a secondary Dirac point (SDP), respectively (Fig. 1(C)). The emergence of the SDP is a consequence of band modulation due to the moiré superlattice resulting from the alignment of graphene and hBN crystals, indicated by the angle $\theta \sim 0°$, which yields energy gaps at the DP and the SDP. With current and voltage terminals of *I*: 62, *V*: 53, the non-local resistance $R_{nl}$ at zero magnetic field is also shown in Fig. 1(C). Near the energy band gap, the hot spots in the graphene due to the hBN, even without a magnetic field, give rise to the transverse bulk topological current, that is, the VHE, which in turn produces a chemical potential difference between terminals 5 and 3, that is, the inverse VHE. A larger $R_{nl}$ of the same order as $h/2e^2$ (= 12.9 kΩ, where *h* is the Planck constant and *e* is the elementary charge) is observed at the SDP. Such a large $R_{nl}$ cannot be explained by the ohmic contribution assuming diffusive transport, with a maximum of 1.5 Ω (see the Supplementary Materials). In contrast, the value of $R_{nl}$ at the DP of approximately 1 kΩ is of the same order as that observed in the previous work (*11*), and the non-local transport at the DP is basically consistent with previous results, implying that bulk valley Hall current emerges and results in non-local transport in our device (*11*). Furthermore, essentially the same phenomena occur at the SDP in the high-temperature regime (see the Supplementary Materials), which implies bulk topological current. However, in the low-temperature regime, the value of $R_{nl}$ at the SDP is even larger than that of $\rho_{xx}$. In this regime, $R_{nl}$ is within the quantum limit, i.e., it equals the quantized value $h/2e^2$ (13) apart from a prefactor of order 1. Furthermore, our results indicate that the $R_{nl}$ itself takes a quantum-limited value and exhibits rigidity under certain conditions, such as a magnetic field, as discussed below. We note that there is an influence of thermal cycles and charge impurities (see the Supplementary Materials). In particular, starting gate voltage fixes the disorder due to charge impurities. The large disorder effects veil the giant non-local resistance and that is what happened in previous studies (11).

Figure 1(D) shows a logarithmic-scale intensity map of the longitudinal conductivity ($\sigma_{xx}$) as a function of $V_g$ and a magnetic field $B$ (applied perpendicular to the substrate). The QHE of single-layer graphene is observed near the DP, and a Landau-fan diagram of Hofstadter's butterfly is observed. From the periodicity of $1/B$, we estimated the alignment angle between the graphene and hBN, $\theta = 0.7$–$0.8°$, and we estimated the moiré superlattice size to be approximately 11 nm. The mobilities in the sample were up to 250,000 cm$^2$/Vs near the DP, and 180,000 cm$^2$/Vs near the SDP at 1.5 K ( see the Supplementary Materials). The mean free path $L_f$ was estimated to be 1–2 μm, which is comparable to the sample geometry.

In small magnetic fields, transverse resistance oscillation was observed, as shown in Fig. 1(E); this was considered to originate from the so-called transverse magnetic focusing (TMF) effect (19). In a ballistic sample, the trajectory of a charged carrier is bent by Lorentz force under a magnetic field, and an injected carrier is focused by the cyclotron motion on the transverse electrode, causing the TMF effect. In our sample, the distance between the electrodes was estimated to be 2.8 μm, based on $B_f = (2\hbar k_F/eL_{eff})i = (2\hbar(\pi n)^{1/2}/eL_{eff})i$, where $B_f$ is the focusing magnetic field, $\hbar$ is the reduced Planck constant given by $h/2\pi$, $k_F$ is the Fermi wave number, $L_{eff}$ is the effective length between electrodes, $n$ is the carrier density, and $i$ is the number of reflections at the edge; this estimated distance is consistent with the nominal center-to-center length of 2.5 μm. The TMF is observed only when the carriers are not scattered during traveling from the injection electrode 6 to the detection electrode 5, thereby achieving the ballistic transport between the electrodes. All of these results confirm the ballistic character of our sample (see the Supplementary Materials).

The experimental energy gap was estimated from the Arrhenius plots: $1/\rho_{xx} \propto \exp(-E_g/2 k_BT)$, where $E_g$ is the gap energy ( = 2Δ), and $k_B$ is Boltzmann's constant, as shown in Fig. 1(F). The gaps were estimated as $E_g = 2\Delta$, being 32 meV and 14 meV at the DP and SDP, respectively, by linear fitting in the high temperature region. The gap at the SDP is consistent with that predicted in (18), although the gap at the DP is much larger than the predicted value of 2 meV. Large gap openings of approximately 30 meV at the DP have been experimentally observed in commensurate graphene/hBN superlattices (20), which is consistent with our result. However, this point remains controversial, and for example, more recent data from angle-resolved photoemission spectroscopy has been reported to support different results (21). We consider that this point is closely connected to the crossover conditions of the transport picture discussed below. A comment is in order on the different scenarios between the DP and SDP. As discussed below, giant non-local resistance occurs only at the SDP in our sample. A hypothesis is that it is attributed to the larger energy gap in the SDP. However, this point is controversial and contradicts our Arrhenius plots. It is also possible that the gap by the Arrhenius plot is different from the energy gap in the band structure. Because the DP shows a longer mean free path than the SDP, another scenario is that ballistic modes can contribute a negative non-local resistance as discussed in ref. (*22*). In contrast to the high-temperature observations, in the lower temperature region, the change of the slope from the Arrhenius form indicates that the carrier transport is dominated by variable range hopping inside the gap due to charge inhomogeneity.

We then investigated the resistances with different terminal configurations in our six-terminal device, and all of the quantum-limited resistance results consistently support the emergence of the quantum valley Hall state (qVHS). In the QHE, two different pictures have been proposed, which are edge picture versus bulk picture. Although it is unsettled which is valid, it is notable that each gives internally consistent results. In our system, the total Hall conductance, indicated by the topological number (*5,6*), vanishes, and there is no a priori reason for edge states to exist. However, recent experimental and theoretical results support the existence of edge states, although their detailed character is determined by their fine electronic structures, which remain to be resolved (*23, 24*). A mechanism driven by the edge states predicts the giant non-local resistance in the quantum limit. Therefore, we calculated the resistances of different configurations using the Landauer-Bütticker formalism with multiterminal samples (*25, 26*), based on a minimal model in which two edge states for two valleys propagate in opposite directions along the edge, with only nonvanishing elements of the transmission matrix (*T*) between terminals given by $T_{j+1,j} = T_{j,j+1} = 1$, where $j$ is an integer, which corresponds to the terminal numbers. The current-voltage relationship is given by $I_k = e^2/h \ \Sigma_j \ (T_{jk}V_k - T_{kj}V_j)$, where $I_k$ is the current flowing out of the $k$-th electrode, $V_j$ is the voltage on the $j$-th electrode, and $T_{jk}$ is the transmission probability from the $j$-th to $k$-th electrode. Solving this equation with an experimental input on $I$ and $V$, we get *I-V* relations and resistances. Figure 2 shows the resistances with different terminal configurations of a six-terminal sample. On the basis of the minimal model, the longitudinal resistance $R_{xx}$ of (*I*: 14, *V*: 65) was calculated as $h/2e^2$, as shown by the dotted lines in Fig. 2(B). The theoretical $R_{nl}$ of (*I*: 61, *V*: 53), (*I*: 61, *V*: 54), and (*I*: 61, *V*: 43) were calculated to be 1/3, 1/6, and 1/6 (unit: $h/e^2$), respectively, as shown by the dotted lines in Fig. 2(C). Similarly, the $R_{nl}$ of (*I*: 62, *V*: 53) and (*I*: 62, *V*: 43) were 2/3 and 1/3 ($h/e^2$), respectively, as shown by the dotted lines in Fig. 2(D). Although the discrepancies between the theoretical predictions may originate from charge impurities (see the Supplementary Materials) and/or the intervalley scattering on the disordered edge with possible dephasing, each $R_{nl}$ is consistent with the theoretical value apart from the fluctuation. Furthermore, the spiking behavior of the resistance is reproducible, and we attribute this effect to mesoscopic fluctuation, which can be considered as a fingerprint of our device and gives an origin of the departure from the quantized theoretical values. Their peaks are classified into two values in Fig. 2 (C and D), and the ratio of the two experimental $R_{nl}$ values is close to 1/2, which implies the existence of channels as proposed in the minimal edge-state model. Overall, all the longitudinal resistances and resistances with nonlocal geometries show a consistent picture. This is in analogy with the QHE, i.e., qVHS in our device. Although all of the $R_{nl}$ peak shapes were asymmetric and cubic scaling did not fit them at low temperatures, the peak shapes became smooth and fit cubic scaling at higher temperatures (see the Supplementary Materials) (*11*), part of which is also suggested by recent theoretical simulations (*24*).

Because the qVHS occurs in zero magnetic field with time reversal symmetry due to the presence of hot spots, its collapse should be connected to the reconstruction of the electronic structure due to a magnetic field through the valley Zeeman energy (VZE), $E_{VZ} = 1/2g^*\mu_B B$, where $g^*$ is an effective *g*-factor and $\mu_B$ is the Bohr magneton. The $g^*$ of a single valley is theoretically given in (*27*) by $m^* = \Delta/v^2, g^* = 2m/m^*$, where $m^*$ is the effective mass, $m$ is the free-electron mass, and $v$ is the Fermi velocity. In our sample, the $g^*$ was theoretically calculated to be approximately 2500 by using the following parameters: $\Delta$ = 7 meV, which is obtained by the Arrhenius plots of the SDP in Fig. 1(F); and $v$ = 1.2 × 10$^6$ m/s, which is determined by Shubnikov-de Haas (SdH) oscillations ( see the Supplementary Materials). This $g^*$ value of the pseudo-spin is 1250 times larger than that of a free electron's spin. Figure 3(B) maps the $R_{nl}$ as a function of $V_g$ and $B$, showing a robust plateau around the zero magnetic field from $B$ = −0.1 to +0.1 T. This result indicates a rigid qVHS when $2\Delta > 2E_{VZ}$ (Fig. 3(A)). After the breakdown of the plateau when $2\Delta < 2E_{VZ}$ (Fig. 3(A)), $R_{nl}$ decreases with increasing magnetic field, approaching zero around 0.8 T (Fig. 3(B) and (C)), and starting to increase in $B$ > 0.8 T due to the occurrence of the quantum Hall states (Fig.3(B)) . The breakdown of the plateau indicates that non-local transport is no longer possible because of the breakdown of the hot spots. Actually, the value of VZE at 0.1 T with $g^*$ = 2500 is $E_{VZ}$ = 7 meV, which agrees well with the result for $\Delta$ = 7 meV of the SDP. Because the model presented in (*27*) is minimal, a prefactor of order 1 can occur in the valley Zeeman term, which is fixed by the details of the electronic structure at the SDP. In this scenario, in the low magnetic field regime, we assume that Landau levels are broadened/overlapped because of thermal effects and disorder. Actually, around the collapse of the plateau, no Landau level is observed and the SdH signal cannot be detected (see the Supplementary Materials). The clear Landau fan is observed only above near 0.5 T in our sample, as shown in Fig. 1(D). This non-local magnetoresistance behavior contrasts with the previously reported simple monotonic increase in $R_{nl}$ (*11, 28*). Our sample, which exhibited lower mobility in several thermal cycles and showed $R_{nl}$ of ~ 1 kΩ, did not show such a clear decrease in the magnetic fields (see the Supplementary Materials), and only samples within the quantum limit with high $R_{nl}$ showed the same behavior. In contrast, $R_{nl}$ in a high-magnetic field regime exhibit a butterfly pattern (see the Supplementary Materials), which should have a QHE-origin, and the SdH-type signal does not occur in the low-magnetic field regime (*29*). As has been previously reported (*28*), with a high magnetic field, spin and energy flow should also affect $R_{nl}$ ( see the Supplementary Materials).

In summary, we measured non-local transport of the ballistic graphene/hBN aligned superlattices with one-dimensional edge contacts. Giant $R_{nl}$ with the order of quantum resistance was observed even at zero magnetic field, indicating the occurrence of the qVHS. Therefore we conclude that the mechanism driven by the edge states is a more likely scenario for the giant non-local resistance in the quantum limit, than a bulk-related interpretation. Furthermore, the $R_{nl}$ plateau, which implies a rigidity of the qVHS, emerged as a function of magnetic field, and its collapse was also observed, which is considered to be a manifestation of valley/pseudospin magnetism. Such an unconventional magnetism should have a potential for engineering the energy-band structure even with a weak magnetic field as well as for spintronics applications.

**Materials and Methods:**
Sample fabrication

First, hBN flakes were exfoliated by applying Scotch tape to a polymethyl methacrylate (PMMA)/polyacrylic acid (PAA)/Si substrate, and then the substrate was floated on a water surface to isolate the PMMA layer from the Si substrate by dissolving the PAA layer in water. The 16-nm-thick hBN flake was then aligned with a graphene flake exfoliated onto a Si substrate with 90-nm-thick SiO$_2$ using a homemade atomic layer transfer system (*30*). The graphene and hBN were carefully aligned by their long edges, and the graphene remained in contact with the hBN flake, forming a graphene/hBN stack. This stack was then aligned with a 20-nm-thick hBN flake that also remained attached to the stack. After removing the PMMA with acetone, we obtained an hBN/graphene/hBN stack on a SiO$_2$/Si substrate. The sample was annealed at 300°C for 30 min in an Ar/H$_2$ atmosphere before atomic force microscopy (AFM) imaging. The AFM image is shown in Supplementary Materials (Fig.S1). The Hall bar geometries are indicated by the black lines. We carefully defined the Hall bar geometry so as to include as few bubbles as possible in the channel region, and the sample was then etched into the Hall bar geometry by reactive ion etching using SF$_6$. The one-dimensional Cr/Au contacts (*31*) were deposited by electron-beam (EB) evaporation followed by EB lithography with a 125 kV acceleration voltage.

Measurement setup

All of the measurements were performed using both four-terminal dc and low-frequency lock-in techniques (around 17 Hz and ac excitation current of 1–10 nA), both of which give consistent results, and measurements were performed in variable temperature cryostats (two types of cryostats were used: base temperatures were 5 and 1.5 K, respectively) with superconducting magnets. All of the experimental data presented in this paper were obtained from the same sample with different cooling cycles from room temperature to below 10 K.

**Acknowledgments:**

We thank Hirotaka Osato of NIMS for technical support of the fabrication process, Katsunori Wakabayashi of Kwansei Gakuin University, Mikito Koshino of Osaka University, Branislav K. Nikolić of the University of Delaware, Sergei Sharapov of the Bogolyubov Institute for Theoretical Physics, and Shu Nakaharai of NIMS for useful comments and discussions.

**Funding:**

The device fabrication and measurement were supported by the Japan Society for Promotion of Science (JSPS) KAKENHI 15K18058, 26630139, and 25706030; and the NIMS Nanofabrication Platform Project, the World Premier International Research Center Initiative on Materials Nanoarchitectonics, sponsored by the Ministry of Education, Culture, Sports, Science, and Technology (MEXT), Japan. Growth of hexagonal boron nitride was supported by the Elemental Strategy Initiative conducted by the MEXT, Japan and JSPS KAKENHI 15K21722.

**Author contributions:**

K.K and S.M. conceived and designed the experiments. K.K., E.W., D.T. fabricated the devices. K.K. and S.M. performed the experiments. K.K., Y.M. and S.M. analyzed the data, developed the models, and wrote the paper. K.W. and T.T. provided the hBN crystals used in the devices.

**Competing interests:**

The authors declare that they have no competing interests.

**Data and materials availability:**

All data needed to evaluate the conclusions in the paper are present in the paper and/or the

Supplementary Materials. Additional data related to this paper may be requested from the authors.

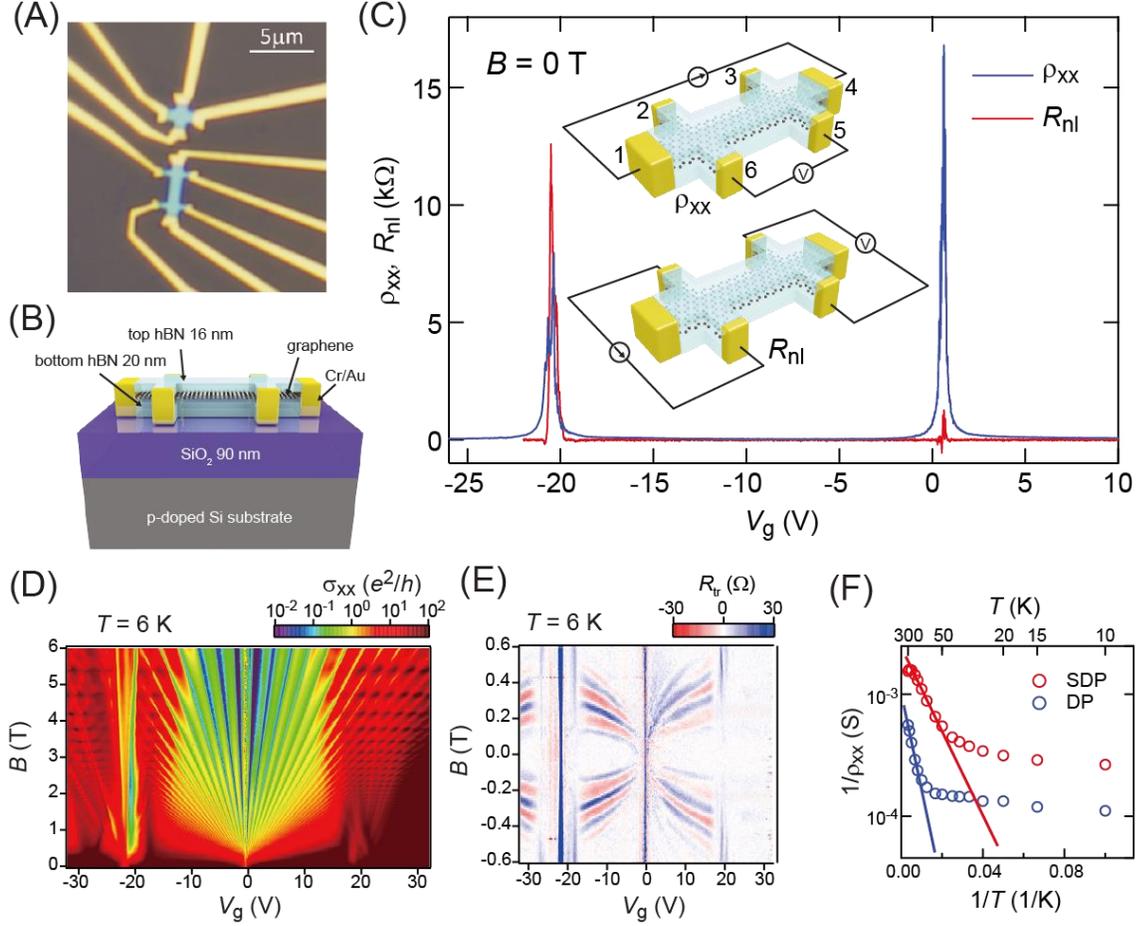

**Fig. 1. Device structure and characterization of our hBN/graphene/hBN superlattices.**

(**A**) Microscope image of our typical hBN/graphene/hBN superlattice device. The Hall bar geometry is such that width $W = 1.0$ μm, length $L = 2.5$ μm. (**B**) Schematic cross-section of the device with Cr/Au contacts on the edge yielding one-dimensional contacts. The backgate voltage ($V_g$) is applied through the 90 nm thickness of $SiO_2$ and the 20 nm thickness of hBN. (**C**) (red curve) Measured non-local resistance $R_{nl}$ (*I*: 62, *V*: 53) and (blue curve) longitudinal resistivity $\rho_{xx}$ (*I*: 14, *V*: 65) as a function of $V_g$ without magnetic fields at 1.5 K. Sharp increases of the $\rho_{xx}$ at the $V_g$ of approximately 0 V and -21 V correspond to a Dirac point (DP) and a secondary Dirac point (SDP), respectively. Inset shows schematic pictures of the measurement setup. (**D**) A logarithmic-scale plot of the longitudinal conductivity ($\sigma_{xx}$) as a function of $V_g$ and magnetic fields (*B*) applied perpendicular to the substrate at 6 K. (**E**) Transverse resistance ($R_{tr}$) oscillation at 6 K, which is $R_{tr}$ as a function of $V_g$ and *B*. The terminal configurations are the same for the $R_{nl}$ in Fig. 1(C). (**F**) Experimental estimation of the energy gaps derived by the Arrhenius plots of the resistivity of the DP and SDP as a function of measured temperature *T*. The gaps are estimated as $E_g = 2\Delta = 32$ meV and 14 meV at the DP and SDP, respectively.

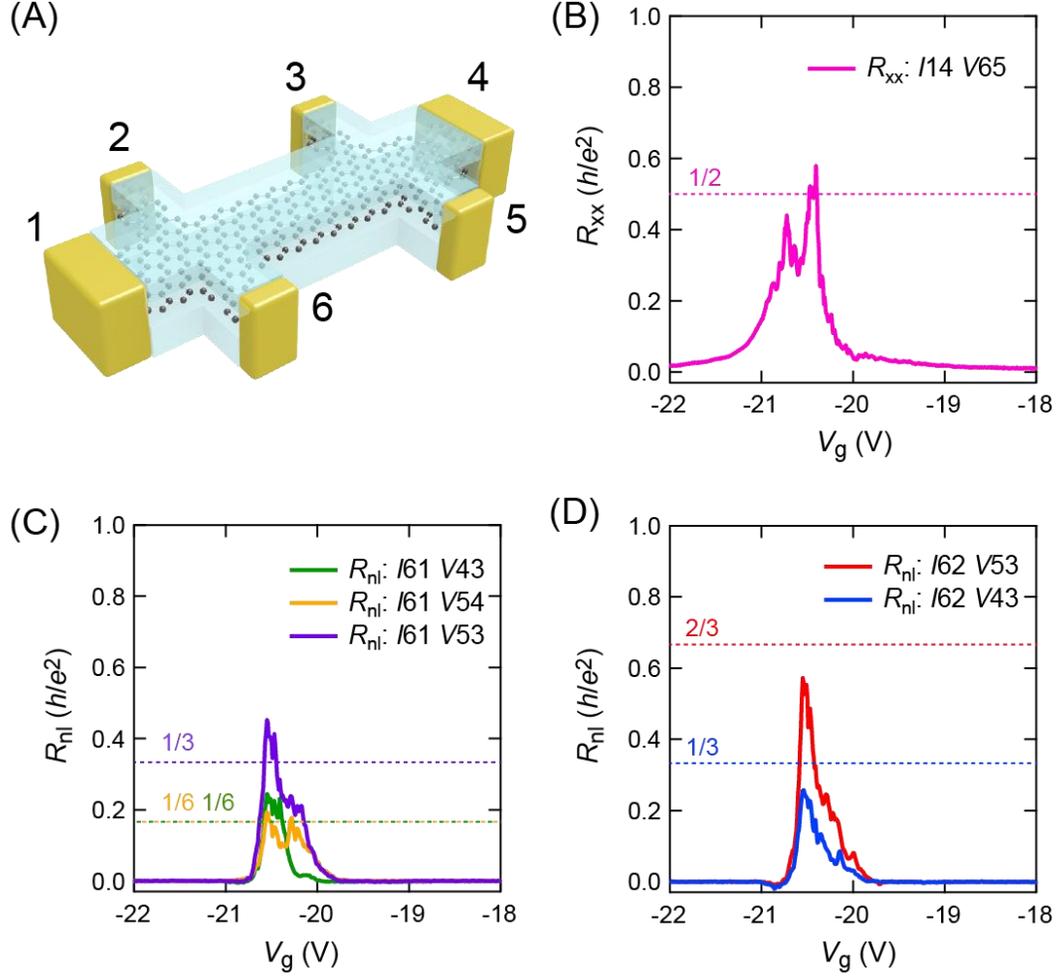

**Fig. 2. Longitudinal (local) resistance $R_{xx}$ and non-local resistance $R_{nl}$ with different terminal configurations on the six-terminal device measured without magnetic fields at 1.5 K.**
(**A**) Schematic showing the terminal number. (**B**) $R_{xx}$ for ($I$: 14, $V$: 65); (**C**) $R_{nl}$ for ($I$: 61, $V$: 53), ($I$: 61, $V$: 54), ($I$: 61, $V$: 43); and (**D**) $R_{nl}$ for ($I$: 62, $V$: 43), ($I$: 62, $V$: 53), as a function of $V_g$. Dotted lines show theoretical values of the resistance based on the minimal edge-state model described in the text. Both $R_{xx}$ and $R_{nl}$ are consistent with the theoretical value apart from fluctuations. The ratio of experimental values of $R_{nl}(I$: 61, $V$: 53)/$R_{nl}(I$: 61, $V$: 54), $R_{nl}(I$: 61, $V$: 53)/$R_{nl}$ ($I$: 61, $V$: 43), and $R_{nl}(I$: 62, $V$: 53)/$R_{nl}(I$: 62, $V$: 43) are approximately 2.

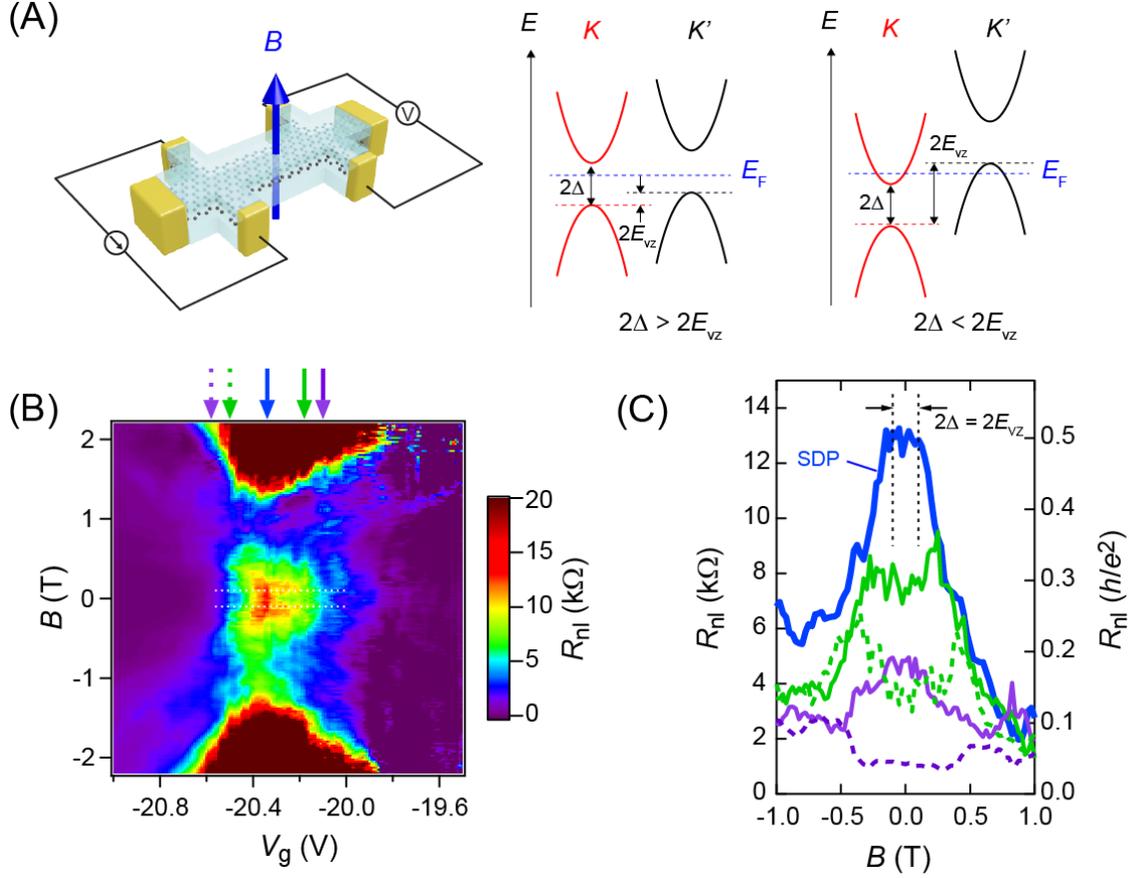

**Fig. 3. Non-local magnetoresistance in graphene superlattices.**
(**A**) Schematics of the measurement set up (left) and the energy band structure in magnetic fields (right), which shows how the band structure is reconstructed when the magnetic field is included. In the band structure, we do not take into account the role of orbital magnetism for simplicity, which leads to broadened Landau bands overlapped due to disorder and finite-temperature effects. (**B**) $R_{nl}$ (*I*: 62, *V*:53) is mapped as a function of $V_g$ and $B$ at 1.5 K. Dashed white lines correspond to the plateau denoted by the black arrows in (**C**). (**C**) $R_{nl}$ vs. $B$ for five $V_g$'s near the SDP, marked with the same color as arrows on the top of (**B**). Black arrows show the regime where energy gap with hot spots is kept, linking to the white dashed lines in (**B**).

# Supplementary Materials for
## Observation of the quantum valley Hall state
## in ballistic graphene superlattices

Katsuyosih Komatsu[*], Yoshifumi Morita, Eiichiro Watanabe, Daiju Tsuya, Kenji Watanabe, Takashi Taniguchi, and Satoshi Moriyama[*]

[*]Correspondence to:  katsuyoshi1.komatsu@toshiba.co.jp, MORIYAMA.Satoshi@nims.go.jp

**This PDF file includes:**

Supplementary Text
  Atomic force microscopy (AFM) image of our sample
  Influence of thermal cycles, and charge impurities
  Experimental estimation of Fermi velocity
  Ballistic character of our device: magnetoresistance due to boundary scattering
  Temperature dependence of the non-local resistance $R_{nl}$ and scaling
  Magnetic field dependence of the non-local resistance $R_{nl}$
Figs. S1 to S6

**Supplementary Text**

Atomic force microscopy (AFM) image of our sample

Figure S1 shows a false-color AFM image of the 16-nm-thick top hBN layer and monolayer graphene after annealing at 300°C for 30 min. The thermal annealing at 300°C may also align the layers.

Influence of thermal cycles and charge impurities

Thermal cycles affect our device's characteristics, including carrier mobility, quantum Hall effect (QHE), and non-local transport. The Landau fan diagram (*15-17*) shown in Fig. 1(D) in the main text was observed after the third cooling cycle to 5 K. Figures S2(A) and (B) show the Landau fan diagrams after the first and fourth cooling cycles to 5 K, respectively. The mobility (and mean free path) was 100,000 cm$^2$/Vs (0.8 μm) near the Dirac point (DP), and 80,000 cm$^2$/Vs (0.6 μm) near the secondary Dirac point (SDP) after the first cooling cycle (Fig. S2(A)), obtained from the Hall effect measurement with low magnetic fields ($B < 0.4$ T). Similarly, the mobility (and mean free path) was 240,000 cm$^2$/Vs (2.2 μm) near the DP and 180,000 cm$^2$/Vs (1.1 μm) near the SDP after the fourth cooling cycle (Fig. S2(B)). For each measurement, the Hofstadter's butterfly patterns were clearly observed, as exemplified by Fig. 1(D) in the main text. However, the QHE of the filling factor of $\nu = 1$ could not be observed in Fig. S2(A) and Fig. S2(B), whereas it was visible for the third cooling cycle, shown in Fig. 1(D). Although the mobility at a low temperature was changed from 100,000 cm$^2$/Vs to 250,000 cm$^2$/Vs near the DP and 50,000 to 200,000 near the SDP after several thermal cycles, the overall pattern of the Landau fan diagram was maintained in every cooling cycle. However, fine structures in the Landau fan were observed to depend on the each cooling process. Figure S2(C) and (D) show the Landau fan diagrams at 80 K and 160 K after the second cooling down, respectively. The Hofstadter's butterfly pattern and the QHE at filling factors of $\nu = 2$ and -2 near the DP were observable up to 160 K. With regard to non-local transport characteristics, the observed non-local resistance $R_{nl}$ near the SDP was ~ 300 Ω in the case with a carrier mobility of 50,000 cm$^2$/Vs, whereas that with a carrier mobility of 180,000 cm$^2$/Vs was ~ 10 kΩ with the same gate-voltage sweeping process and the same measurement configuration at 5 K. We attribute the mobility shift to the modulation of the charge inhomogeneity that occurred after several thermal cycles, which may influence the transport property, and in particular, the $R_{nl}$ value (*11*). In order to obtain a large $R_{nl}$ and a large valley Hall effect, high-quality samples are required, and therefore we also address the influence of charge impurities on our device. Figure S2(E) shows the $R_{nl}$ with various starting gate voltages $V_{g,start}$ at the fourth cooling cycle to 5 K. $V_g$ was first swept from 0 V to $V_{g,\,start}$, and then $R_{nl}$ was measured from $V_{g,\,start}$ to 0 V. The right inset in Fig. S2(E) shows that larger shifts in peak positions occurred with increasing $V_{g,\,start}$, whereas the peak positions approached a constant value near $V_g = -21$ V with decreasing $V_{g,\,start}$, indicating the occurrence of a charge accumulation at the interface between hBN and SiO$_2$, as well as an increasing influence of charge impurities with a higher $V_{g,\,start}$. The charge impurity may cause disorder in the channel, which would increase the scattering of carriers. Therefore, as shown in the left inset in Fig. S2(E), larger $R_{nl}$ for lower $V_{g,\,start}$ was observed because of the reduced influence of charge impurity, whereas $\rho_{xx}$ became smaller for lower $V_{g,start}$. Such $R_{nl}$ behavior due to charge inhomogeneity is consistent with that previous reported (*11*). In contrast, we note that the overall pattern of the Landau fan diagram is robust with respect to initial gate voltage.

Experimental estimation of the Fermi velocity

Figure S3(A) shows the magnetic field dependence of the longitudinal conductance at temperatures from 5 K to 60 K after the fourth cooling cycle. The conductance oscillation is a so-called Shuvnikov-de Haas oscillation (SdHO). Fig. S3(B) shows the SdHO amplitude, defined as $\Delta G \propto 2\pi^2 k_B T/\sinh(2\pi^2 k_B T m_c/\hbar eB)$, where $m_c = \hbar (\pi n)^{1/2}/v$. Fitting to this formula yields the Fermi velocity $v = 1.2 \times 10^6$ m/s for $n$ of approximately $10^{11}$ cm$^{-2}$, which is approximately 20% larger than the graphene on top of an oxidized silicon wafer, $v = 1 \times 10^6$ m/s (*32*). This increase in $v$ is explained by electron-electron interactions in the low carrier density region, where $n$ is approximately $10^{11}$ cm$^{-2}$ (*33*).

Ballistic character of our device: magnetoresistance due to boundary scattering

Figure 1 of the main text establishes ballistic character of our sample. Here we further substantiate its ballistic character, based on magnetoresistance due to boundary scattering. This phenomenon has not been reported in graphene on SiO$_2$, even in a suspended case. In conventional mesoscopic 2DEG, an anomalous peak structure in the magnetotransport is well-known to occur because of the commensurability between cyclotron motion and sample geometry (*34*). Fig. S4 shows the magnetotransport in our ballistic hBN/graphene/hBN sample, where the mean free path can be seen to be comparable to the sample geometry, and has characteristics of peak structure due to boundary scattering in a low-magnetic field regime (*35*) from low temperature to moderately high temperature, e.g., $T = 70$ K. These results further support the ballistic character of our devices.

Temperature dependence of $R_{nl}$ and scaling

In our paper, we employed a Hall bar geometry to observe $R_{nl}$. Non-local voltage is a difference in potential that occurs away from the nominal current path. There are at least two origins for non-local voltage in our sample: the mediated bulk topological valley current and diffusive transport. Concerning the former, a cubic scaling formula ($R_{nl} \propto \sigma_{VH}^2 \rho_{xx}^3$, where $\sigma_{VH}$ is the valley Hall conductance) has been proposed (*36*, *37*) and was also confirmed experimentally (*11*, *38-40*). We found that although the cubic scaling holds in the high-temperature regime at the SDP (Fig. S5 (C)), the formula fails in the low-temperature limit for the SDP at both the hole and the electron sides (Fig. S5 (B)), where $R_{nl}$ is close to the quantum-limited value. Figure S5 (A) shows the temperature dependence of the $R_{nl}$ of the SDP, wherein the spiking structures in the $R_{nl}$ at low temperatures can be seen to become smooth at high temperatures. Concerning the DP, we confirmed that our data are consistently described by cubic scaling (*11*). From the temperature dependence of the $R_{nl}$, we also estimated the activation energy derived from the Arrhenius plots, assuming that $1/R_{nl} \propto \exp(-A/2 k_B T)$, where $A$ is the activation energy, as shown in Fig. S5 (D). The experimental activation energies were estimated to be 92 meV and 40 meV at the DP and the SDP, respectively, by linear fitting in the high temperature region. The obtained activation energies are approximately three times larger than the activation energy that was obtained from the local transport (Fig. 1(F) in the main text). However, these results are considered to be reasonable in the case that the cubic scaling ($R_{nl} \propto \sigma_{VH}^2 \rho_{xx}^3$) is established in the linear fitting region (*40*). Regarding diffusive transport, which also leads to non-local voltage, a current injected from the source does not flow along the nominal path but is expected to spread into our sample. This phenomenon is described by $(\rho_{xx}/\pi)\exp(-\pi L/W)$, which is the formula used in our study (*41*), and the estimated maximum contribution is 1.5 Ω, basically negligible in the geometry of the sample presented in this study. Finally, regarding a recent proposal that the non-locality can be induced by Fermi surface edge states, the scenario illustrates the origin of non-locality at the ballistic limit and the failure of cubic scaling as shown above (*24*). We consider that a crossover from a bulk phenomenon to an edge phenomenon occurs at the SDP. The detailed conditions for the crossover are expected to be connected to the electronic structures at the DP and SDP, such as the scale of the energy gap and mobility, which remain to be solved.

Magnetic field dependence of $R_{nl}$

Our manuscript focuses on the low-magnetic field regime of the non-local magnetoresistance, where the superlattice structure in our device plays a main role in the non-local transport. Here, to complement the main text, we show data on the high-magnetic field regime. As shown in Fig. S6(A) (which can be compared with Fig. 1(D) showing the longitudinal conductivity in the main text), a butterfly pattern of the $R_{nl}$ is observed in the high-magnetic field regime, which reflects the zero points of the longitudinal resistivity and should have a QHE origin. The TMF is also observed in the low-magnetic field regime below 0.5 T (*19*). Moreover, as has been reported before (*28*, *42*), bulk spin and heat current should also play a role in $R_{nl}$ under magnetic field and broken time-reversal symmetries. Fig. S6(B) shows the magnetic field dependence of the second harmonic of the excitation frequency in $R_{nl}^{2F}$, focusing on the SDP for simplicity. No signal is visible at 0 T, and the butterfly pattern is also observed under magnetic fields. We consider that the increase of $R_{nl}^{2F}$ is, at least in part, due to combined thermal effects, e.g., Joule heating and Nernst effects, as discussed in (*42*). Furthermore, the rapid increase of the non-locality at the DP and SDP under a magnetic field is expected to be connected to edge-state transport in the QHE regime.

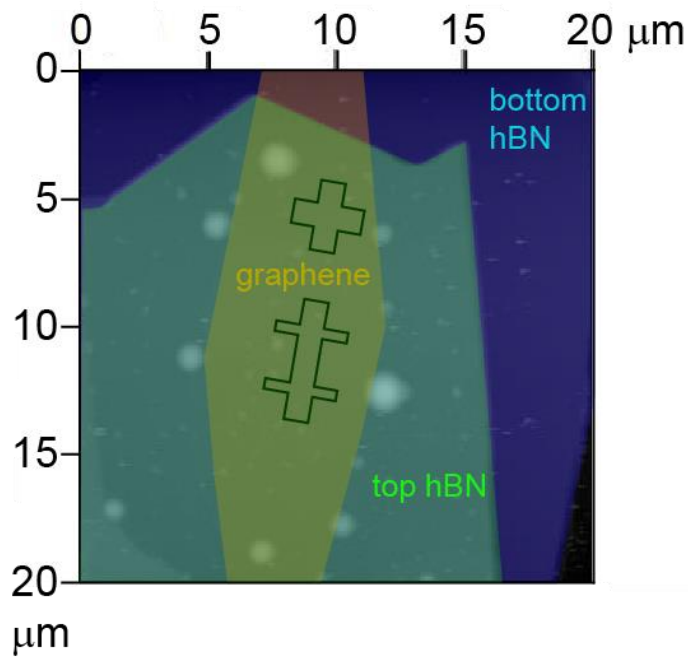

**Figure S1. False-color AFM image of our hBN/graphene/hBN stack.** The top hBN layer is 16-nm-thick, the bottom hBN layer is 20-nm-thick, and the graphene is a monolayer.

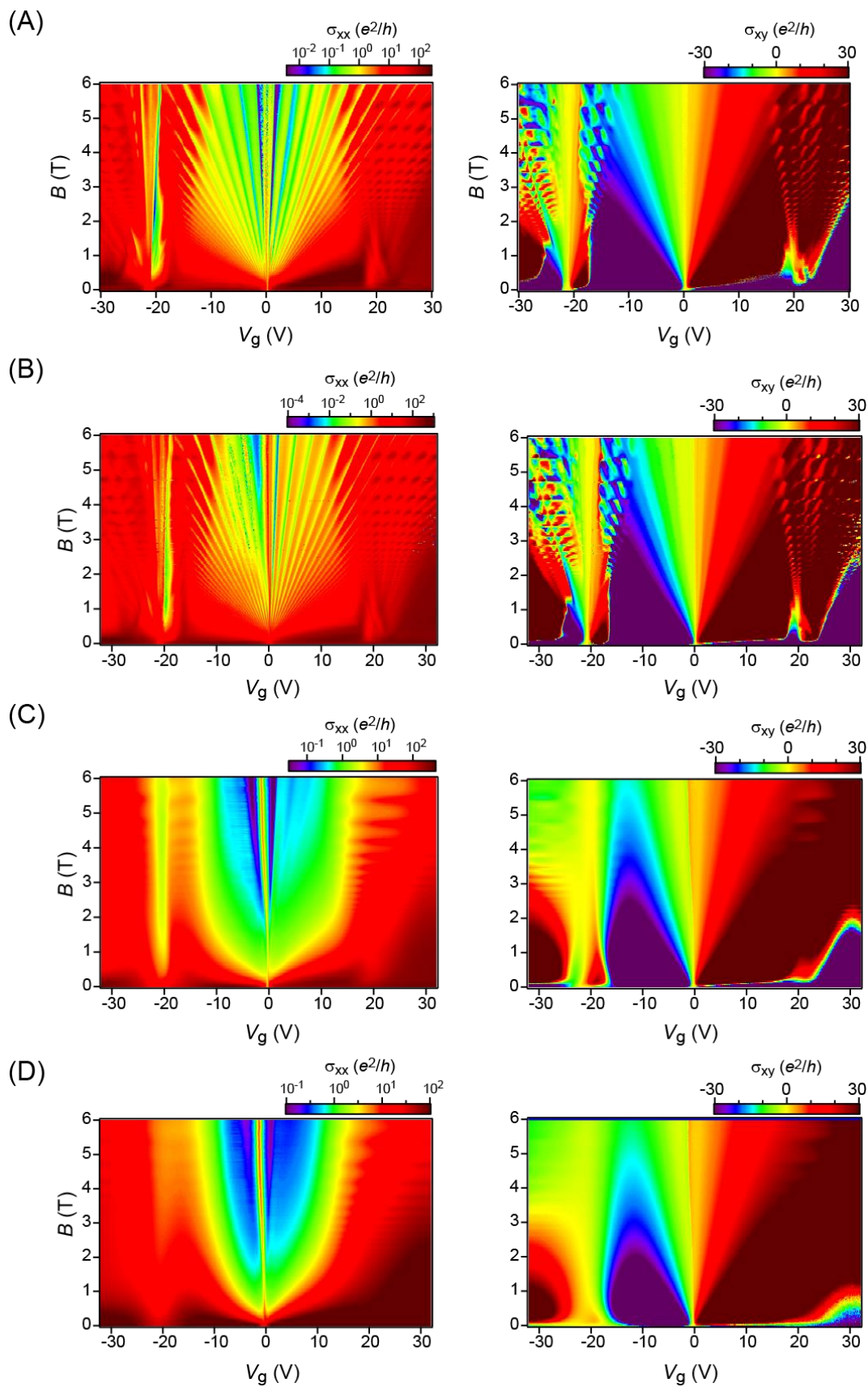
18

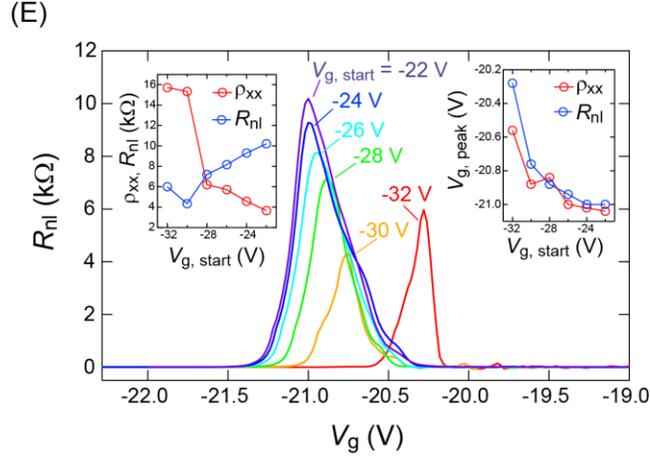

**Figure S2. Influence of thermal cycles and charge impurities on our device.** (A) Logarithmic-scale intensity map of $\sigma_{xx}$ as a function of $V_g$ and magnetic field $B$, and linear-scale intensity map of $\sigma_{xy}$ after the first cooling cycle at 5 K. (B) $\log(\sigma_{xx})$ and $\sigma_{xy}$ after the fourth cooling cycle at 5 K. (C) $\log(\sigma_{xx})$ and $\sigma_{xy}$ at 80 K after the second cooling down. (D) $\log(\sigma_{xx})$ and $\sigma_{xy}$ at 160 K after the second cooling down. (E) $R_{nl}$ with various starting gate voltage $V_{g,\text{start}}$ after the fourth cooling cycle. Left inset shows $V_{g,\text{start}}$ dependence of the peak height. Right inset shows $V_{g,\text{start}}$ dependence of the peak position.

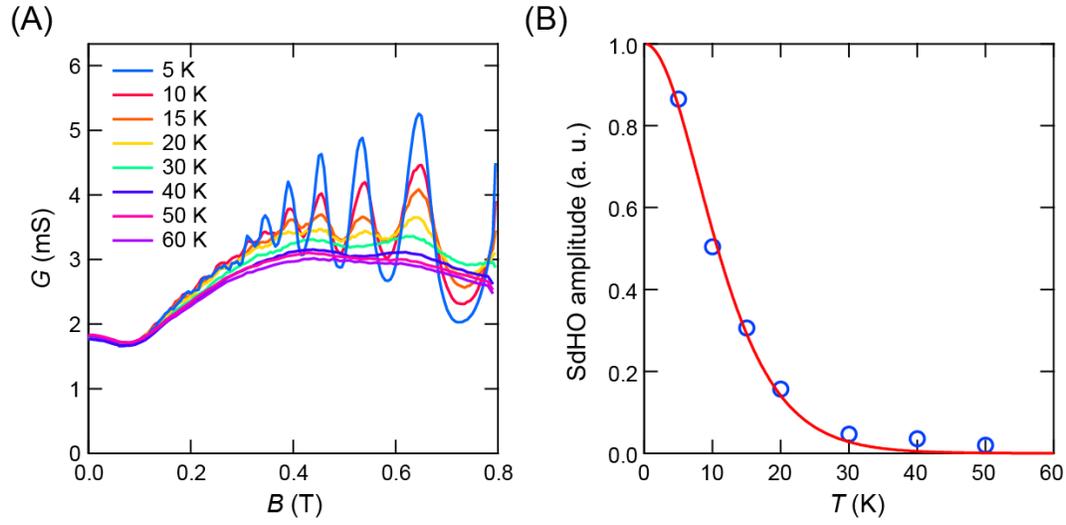

**Figure S3. Shubnikov-de Haas oscillation and estimation of the Fermi velocity.** (A) Shubnikov-de Haas oscillation (SdHO) of our hBN/graphene/hBN device. Conductance is shown as a function of magnetic field. (B) Amplitude of the SdHO in our device as a function of temperature fitted by the SdHO formula.

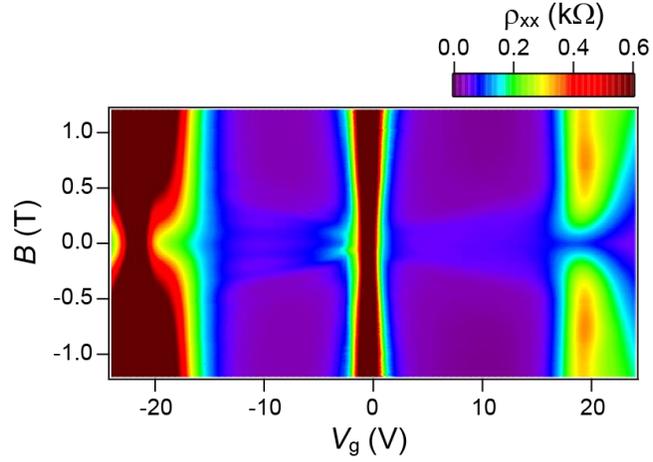

**Figure S4. Magnetotransport in our ballistic hBN/Graphene/hBN sample.** Intensity mapping of $\rho_{xx}$ as a function of $V_g$ and $B$ measured at $T = 70$ K. The color-scale is limited up to 0.6 k$\Omega$ in order to focus on the ballistic characteristics. Transport characteristics of four peak structures as a function of $B$ due to boundary scattering are observed between the DP and the hole-side of the SDP, from -0.5 to 0.5 T. The two peak structures are also fairly visible between the DP and the electron side of the SDP.

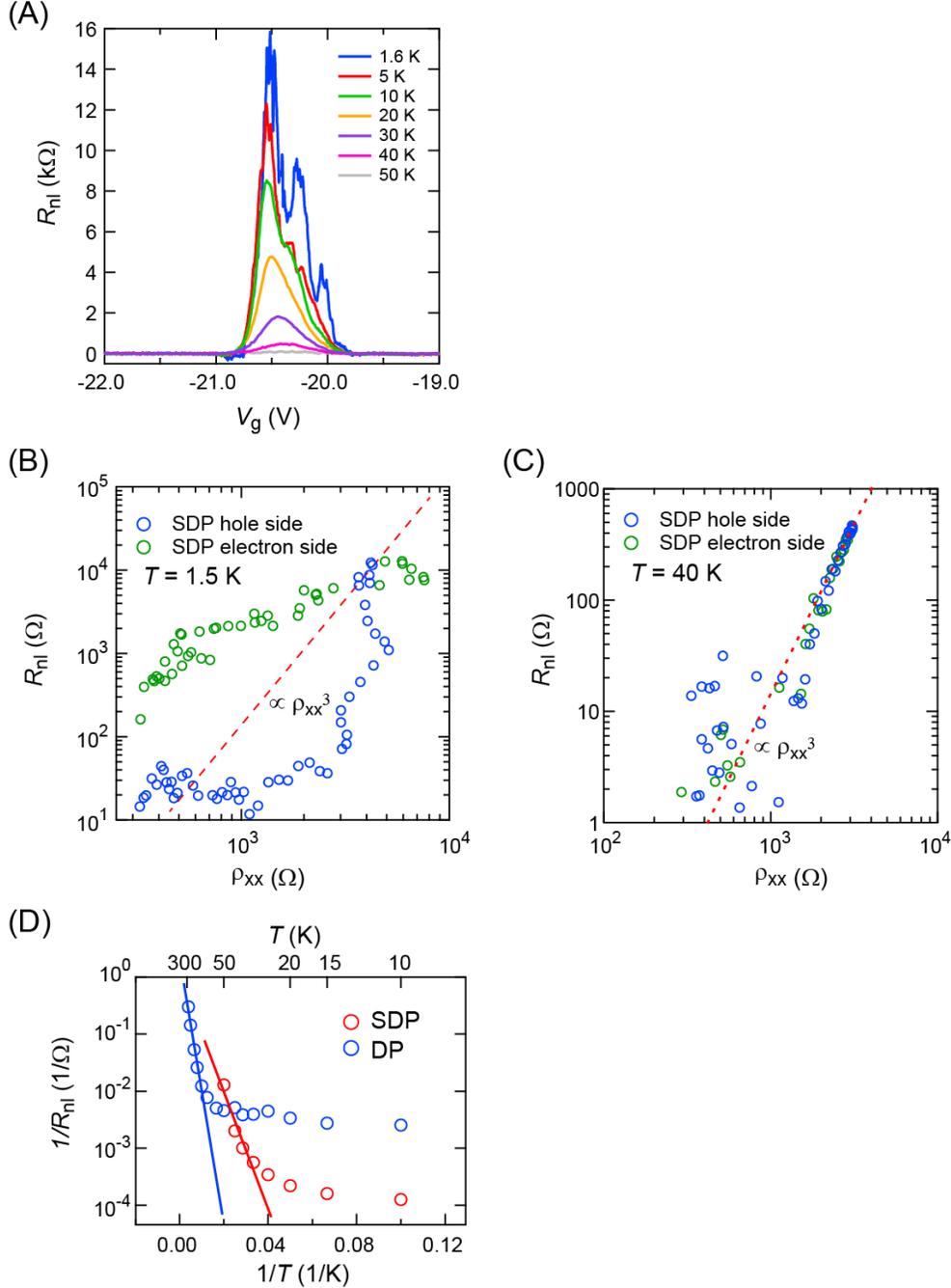

**Figure S5. Temperature dependence of the non-local resistance $R_{nl}$.** (A) Temperature dependence of the $R_{nl}$ at the SDP. (B) Scaling of $R_{nl}$ as a function of local resistivity $\rho_{xx}$ at 1.5 K. (C) Scaling of $R_{nl}$ as a function of $\rho_{xx}$ at 40 K. At the SDP, in the high-temperature regime, the cubic scaling generally holds, which implies bulk topological currents (*11*). In contrast, in the low-temperature regime, a simplistic scaling does not hold. (D) Arrhenius plot of $R_{nl}$ of the DP and SDP. The estimated activation energy is 92 meV for the DP and 40 meV for the SDP, which are both approximately three times larger than the activation energy obtained from the local transport, $E_g = 2\Delta = 32$ meV for the DP and 14 meV for the SDP.

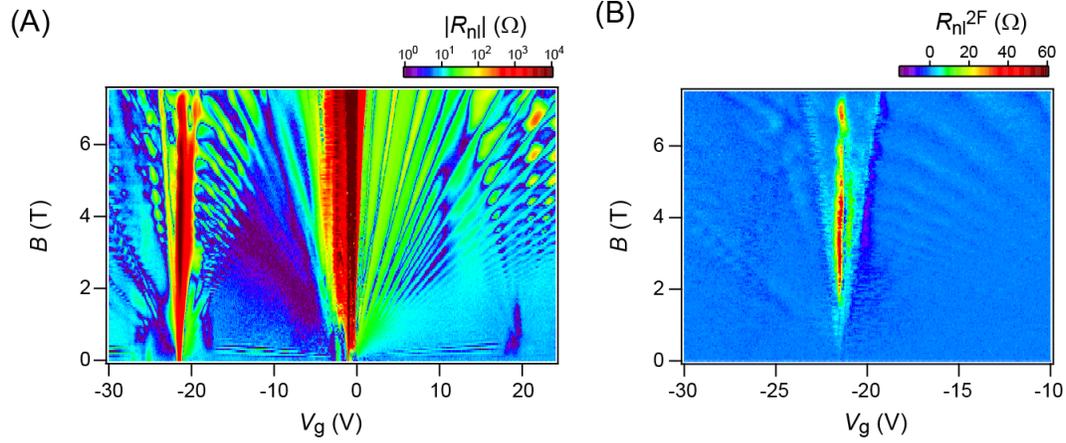

**Figure S6. Magnetic field dependence of the $R_{nl}$.** (A) Hofstadter butterfly in the non-local magnetoresistance of our graphene superlattice. Logarithmic-scale $R_{nl}$ is mapped as a function of $V_g$ and $B$ at 1.5 K. Note that TMF is also observed in the low-magnetic field regime below 0.5 T (*19*). (B) Linear-scale intensity mapping of the second harmonic signal of $R_{nl}^{2F}$ around the SDP at 1.5 K. AC excitation current is 10 nA.